\documentclass[5p,times]{elsarticle}

\usepackage{algpseudocode}
\usepackage{amsmath}
\usepackage{hyperref}

\journal{journal}

\begin{document}

\begin{frontmatter}

\title{Open-source software for electrical engineering applications requiring consideration of electrodynamics: elecode}

\author{Dmitry Kuklin\corref{mycorrespondingauthor}}
\address{Northern Energetics Research Centre, Kola Science Centre of the Russian Academy of Sciences, Apatity, Russia}

\cortext[mycorrespondingauthor]{Corresponding author}
\ead{kuklindima@gmail.com}

\begin{abstract}
The work presents elecode, open-source software for various electrical engineering applications that require considering electromagnetic processes. The primary focus of the software is power engineering applications. However, the software does not impose any specific limitations preventing other uses.
In contrast to other open-source software based on the Finite Difference Time Domain (FDTD) method, elecode implements various thin wire modeling techniques which allow simulating complex objects consisting of wires. In addition, implemented graphical user interface (GUI) helps modify models conveniently.
The software provides auxiliary numerical methods for simulations and measurements of the electrical soil properties, allows conducting lightning-related simulations (including those involving isolation breakdown models), and calculations of grounding characteristics.
The part of the code responsible for FDTD simulations is well tested in previous works. Recently, the code was rewritten in order to add a convenient interface for using it as a library, command-line program, or GUI program. Finally, the code was released under an open-source license.
The main capabilities of the software are described in the work. Several simulation examples covering main software features are presented. elecode is available at https://gitlab.com/dmika/elecode.
\end{abstract}


\end{frontmatter}

\section{Introduction}
Various applications in electrical engineering require considering complex electromagnetic phenomena \cite{grcev_electromagnetic_1990,tanabe_novel_2001,yutthagowith_application_2011,olsen_comparison_1996}. Examples of such applications are simulations of overvoltages caused by lightning strikes \cite{kuklin_choosing_2016}, calculations of electrical grounding characteristics (such as grounding impedance, ground potential rise, etc.) \cite{kuklin_extension_2016}, and simulations related to measurements of frequency-dependent soil properties \cite{kuklin_numerical_2019,kuklin_device_2020}.

In order to meet the necessary requirements, suitable software is needed. FDTD method has demonstrated great potential in solving power engineering problems. However, existing open-source FDTD software is usually designed for other purposes \cite{warren_gprmax_2016,noauthor_meep_2022}. Lack of thin wire techniques, absence of convenient graphical user interface (GUI) for modifying 3D models, and lack of additional numerical methods (for measurements of soil properties, insulation breakdown, etc.) do not allow using the existing software for power engineering applications. Closed-source software does not allow the implementation of new numerical methods and therefore is very limited for research purposes.

To fill this gap, open-source software was developed that allows conducting simulations conveniently both at the level of computer code and user interface level.

The software mainly focuses on power engineering applications but is not limited to them: any other uses are possible (especially if computer code is expanded with additional numerical methods).

Different parts of the code were used previously in various problems \cite{kuklin_choosing_2016,kuklin_extension_2016,kuklin_numerical_2019,kuklin_device_2020}. However, the code was closed source, fragmented, and unsuitable for convenient user-friendly usage. Recently the code has been rewritten in a more convenient for modifications manner, put together as a single software toolset, and released under an open-source license.

\section{Numerical methods}
In this section, the main numerical methods implemented in the software are described. Apart from the FDTD method, numerous additional methods were implemented which can be used in electrical engineering.

\subsection{Basic FDTD algorithm}
The core of the software is the implementation of the FDTD method (and additional FDTD-related techniques).

The basic FDTD algorithm is based on two of the Maxwell equations \cite{yee_numerical_1966,taflove_numerical_1975}:
\begin{equation}\label{eq:maxwell}
\begin{split}
& \frac{\partial\mathbf{B}}{\partial t} = -\mathbf{\nabla}\times\mathbf{E}, \\
& \frac{\partial\mathbf{D}}{\partial t} = \mathbf{\nabla}\times\mathbf{H} - \mathbf{J}.
\end{split}
\end{equation}

In Cartesian coordinates, components of the electric field:

\begin{equation}
\begin{split}
& \frac{\partial E_x}{\partial t} = \frac{1}{\varepsilon} \left[ \left( \frac{\partial H_z}{\partial y} - \frac{\partial H_y}{\partial z} \right) - \sigma E_x \right] \\
& \frac{\partial E_y}{\partial t} = \frac{1}{\varepsilon} \left[ \left( \frac{\partial H_x}{\partial z} - \frac{\partial H_z}{\partial x} \right) - \sigma E_y \right] \\
& \frac{\partial E_z}{\partial t} = \frac{1}{\varepsilon} \left[ \left( \frac{\partial H_y}{\partial x} - \frac{\partial H_x}{\partial y} \right) - \sigma E_z \right],
\end{split}
\label{eq:maxwelle}
\end{equation}
where $\varepsilon$ and $\sigma$ are permittivity and conductivity of the medium.

For the magnetic field:

\begin{equation}
\begin{split}
& \frac{\partial H_x}{\partial t} = \frac{1}{\mu} \left( \frac{\partial E_y}{\partial z} - \frac{\partial E_z}{\partial y} \right) \\
& \frac{\partial H_y}{\partial t} = \frac{1}{\mu} \left( \frac{\partial E_z}{\partial x} - \frac{\partial E_x}{\partial z} \right) \\
& \frac{\partial H_z}{\partial t} = \frac{1}{\mu} \left( \frac{\partial E_x}{\partial y} - \frac{\partial E_y}{\partial x} \right),
\end{split}
\label{eq:maxwellh}
\end{equation}
where $\mu$ is the permeability of the medium.

Yee proposed using finite difference expressions for these equations \cite{yee_numerical_1966}.
According to this approach, the $E_x$ component at a certain time moment can be calculated as \cite{taflove_numerical_1975}:

\begin{equation}
\begin{split}
& {E_x}|^{n+1}_{i+1/2,j,k} = \Bigg( 1 - \frac{\sigma_{i+1/2,j,k} \cdot \Delta t}{\varepsilon_{i+1/2,j,k}} \Bigg) \cdot {E_x}|^{n}_{i+1/2,j,k} \\
& + \frac{\Delta t}{\varepsilon_{i+1/2,j,k}} \cdot \Bigg( \frac{{H_z}|^{n+1/2}_{i+1/2,j+1/2,k} - {H_z}|^{n+1/2}_{i+1/2,j-1/2,k}}{\Delta y} \\
& - \frac{{H_y}|^{n+1/2}_{i+1/2,j,k+1/2} - {H_y}|^{n+1/2}_{i+1/2,j,k-1/2}}{\Delta z} \Bigg). \\
\end{split}
\label{eq:fdtdex}
\end{equation}
Similarly for other electric field components.

$H_x$ magnetic field component is calculated as \cite{taflove_numerical_1975}:

\begin{equation}
\begin{split}
& {H_x}|^{n+1/2}_{i,j+1/2,k+1/2} = {H_x}|^{n-1/2}_{i,j+1/2,k+1/2} \\
& + \frac{\Delta t}{\mu_{i,j+1/2,k+1/2}} \cdot \Bigg( \frac{{E_y}|^{n}_{i,j+1/2,k+1} - {E_y}|^{n}_{i,j+1/2,k}}{\Delta z} \\
& - \frac{{E_z}|^{n}_{i,j+1,k+1/2} - {E_z}|^{n}_{i,j,k+1/2}}{\Delta y} \Bigg).
\end{split}
\label{eq:fdtdhx}
\end{equation}
Analogously for remaining magnetic field components.

Using these equations, it is possible to calculate electric and magnetic fields for a particular simulation volume and a certain number of time steps (assuming that the field source is simulated too).

For the cases when better accuracy, special properties of the media, field sources, or additional numerical techniques are required, these equations are modified accordingly \cite{taflove_computational_2005}.

For simulations, it is necessary to model sources of the electromagnetic field.
The software allows modeling different kinds of sources: non-ideal and ideal current and voltage sources \cite{taflove_computational_2005} (they are used the most for the tasks the software focuses on), pointwise electric and magnetic field sources \cite{taflove_computational_2005}, and plane-wave sources. The plane wave is modeled with the total-field / scattered-field technique \cite{taflove_computational_2005}, and it can be useful for modeling antennas (or testing thin wire techniques).

One of the important objects in electrical engineering tasks are lumped circuit elements. Such elements as resistors and capacitors can be modeled with the software \cite{taflove_computational_2005}.

\subsection{Absorbing boundary conditions}
To simulate the extension of the computation domain to infinity, absorbing boundary conditions (ABC) are required \cite{taflove_computational_2005}.

Among the existing ABCs, the perfectly matched layer (PML) is one of the most efficient. Two types of PML were implemented in the software: uniaxial PML (UPML) \cite{berenger_perfectly_1994,taflove_computational_2005} and convolutional PML (CPML) \cite{roden_convolution_2000,taflove_computational_2005}.

\subsection{Thin wire techniques}
Another essential object that is usually required in simulations is long wire (with a diameter significantly smaller than the cell size). There are multiple thin wire techniques. Some of them are suited for wires located along the FDTD grid. In some cases, arbitrarily located wires are needed. Three main wire models are implemented: the model \cite{railton_treatment_2005} with correction \cite{taniguchi_improved_2008,taniguchi_improved_2009}, staircase wire model \cite{montoya_modeling_1999,noda_error_2004}, and oblique thin wire technique \cite{guiffaut_new_2012}.

\subsection{Dispersive media}
In certain cases, it is necessary to model dielectrics with dispersive properties. One of such cases is modeling frequency-dependent soil properties \cite{akbari_effect_2013,cavka_comparison_2014,alipio_frequency_2013,datsios_methods_2020,visacro_lightning_2015,li_inversion_2020}. For this purpose, the Debye model provides acceptable results (if the required frequency range is not too wide) \cite{kuklin_extension_2016}.

For simulating Debye media, the auxiliary differential equation (ADE) method is used in the software \cite{okoniewski_simple_1997,taflove_computational_2005}.

The thin wire techniques \cite{railton_treatment_2005} and \cite{guiffaut_new_2012} mentioned above are modified for usage with Debye media according to \cite{kuklin_extension_2016}.

\subsection{Additional methods}
Apart from the FDTD-related methods, the software also provides the implementation of additional numerical methods and mathematical models useful for various electrical engineering applications.

In high-voltage engineering, modeling electrical soil properties is essential because electrical grounding characteristics depend on them. Therefore, many different numerical methods were implemented that are necessary for measurements and simulations of the soil properties and grounding characteristics.

Using the Debye model for (simulating dielectric properties) requires calculating the Debye model parameters. For this purpose, the hybrid particle swarm-least squares optimization approach was implemented \cite{kelley_debye_2007}.

For calculating apparent soil properties, the method \cite{takahashi_analysis_1990} was implemented. This method is important for low-frequency soil resistivity measurements (but can be useful for higher frequencies too).

Soil measurements also require information on various characteristics of a measurement array \cite{loke_tutorial_2020}. One of the characteristics, depth of investigation, can be calculated with the implemented methods for dipole-dipole arrays \cite{roy_depth_1971} and \cite{barker_depth_1989}.

Additionally, several models for frequency-dependent soil properties were implemented: Alipio-Visacro model \cite{alipio_modeling_2014}, Portela model \cite{portela_measurement_1999,cavka_comparison_2014}, and Messier model \cite{cavka_comparison_2014}.

There is also additional code implementing insulation breakdown models \cite{shindo_new_1985,pigini_performance_1989}.
The models can be used, for example, for calculating back flashover probability due to lightning strikes \cite{kuklin_choosing_2016,chisholm_new_2010}.
Various implemented lightning current waveforms \cite{heidler_class_2002,noauthor_guide_1991} help conduct calculations related to lightning strikes.

\section{Software}

\subsection{Programming language and libraries}
The code uses the C++ programming language. C++ was chosen because it allows programming both at high and low levels. In addition, it is fast and ubiquitous: GNU Compiler Collection used for the compilation is readily available on many platforms (especially Linux-based ones). There are plenty of useful libraries written for C++.

The code avoids dependencies on external libraries as much as possible to simplify software maintenance and usage. Most of the code only needs open-source GNU Compiler Collection (which includes OpenMP required for parallel computing).
However, additional libraries are needed for the GUI. The GUI uses Qt \cite{noauthor_qt_nodate} and OpenGL \cite{noauthor_opengl_nodate} libraries.

\subsection{Software structure}
\begin{figure}[!t]
\centering
\includegraphics[width=2.7in]{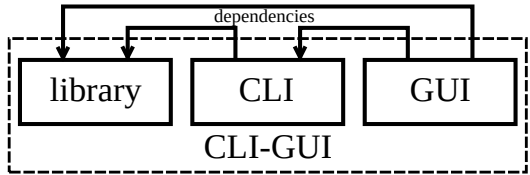}
\caption{Software modules and their dependencies.}
\label{fig:depend}
\end{figure}

The software is divided into several modules (Fig.~\ref{fig:depend}): library, command-line interface (CLI), and GUI. This structure creates the possibility for conducting simulations at different levels and helps avoid the influence of higher level code (such as GUI, for example) on low-level code (responsible for the simulation process). Thanks to this approach, if a GUI level error occurs during the simulation process, the simulations would continue. Also, this helps avoid dependencies on external libraries (Qt, OpenGL) if there is no need for them.

The library allows the creation of custom programs for which relatively low-level usage is required. It provides full control over implemented numerical methods. But using the library is not as convenient as a program with a user interface.

Command-line interface (CLI) is created for running simulations in the terminal according to model files or entered commands. This is important for conducting simulations conveniently when GUI is not required. For example, this allows automating simulations via scripting or using an operating system without a window system.

The primary purpose of GUI is to create, modify, and check models conveniently.

The software can be compiled as a CLI-GUI program which is the default option and most convenient to use.

\subsection{Interface}
CLI has more capabilities for running simulations than GUI: all essential numerical methods implemented in the software are available through CLI.

GUI is mostly used for preparing models for FDTD simulations. If necessary, GUI can launch elecode in CLI mode for running simulations. As stated above, this allows the independent execution of GUI and CLI programs.

\begin{figure}[!t]
\centering
\includegraphics[width=3.4in]{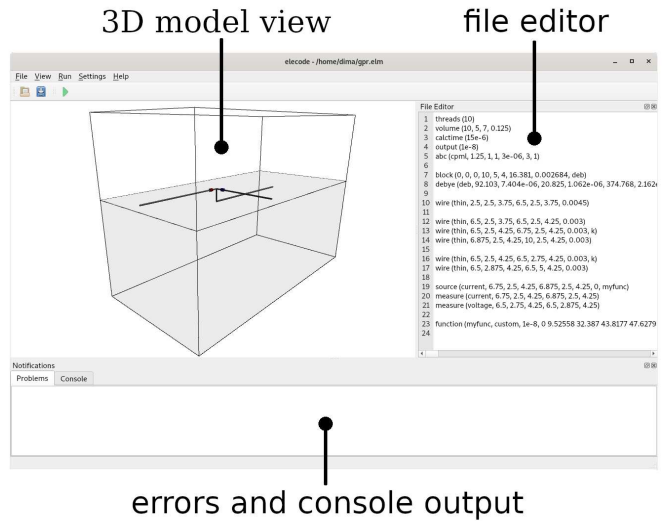}
\caption{Graphical user interface.}
\label{fig:gui}
\end{figure}

The main GUI window is shown in Fig.~\ref{fig:gui}. It includes a 3D model view, an editor for modifying model files (saved in plain text format), an error panel (for warnings and errors in the model file), and console output (for output from CLI).

If compiled as a CLI-GUI application, elecode can be used in CLI and GUI modes. This is convenient because both options are used in practice: GUI for creating models and CLI for running simulations.

\section{Examples of use}
Several examples are presented in this section to demonstrate the main possibilities of the software.

\subsection{Electrical grounding characteristics}
As mentioned above, the software can be used for calculations of electrical grounding characteristics. The previous version of the code has been successfully used for calculations of the ground potential rise \cite{kuklin_extension_2016}. This section presents an example in which measured ground potential rise for buried horizontal wire is compared to that calculated with the software.

Measurements were carried for 4~m long buried wire. After the measurements of the ground potential rise, the buried wire was removed and soil properties (resistivity and permittivity) were measured. Then, after the measurements of the soil properties, the complex permittivity was calculated with the software. Finally, Debye model parameters were calculated with the software. This allowed using measured soil properties in the FDTD simulation. Debye parameters for the measured soil properties are shown in Table~\ref{parammeas}).

\begin{table}[!t]
\renewcommand{\arraystretch}{1.3}
\caption{The $\epsilon_{\infty}$, $\Delta\epsilon$, and $\tau$ parameters of the four-term Debye function expansion.}
\centering
\begin{tabular}{cccc} \hline
$\rho_0$, $\Omega\cdot$m  & 372.58  \\ \hline
$\sigma_0$, mS/m          & 2.684   \\
$\epsilon_\infty$         & 16.381  \\
$\Delta\epsilon_1$        & 92.103  \\
$\tau_1$, s               & 7.404$\cdot 10^{-6}$  \\
$\Delta\epsilon_2$        & 20.825  \\
$\tau_2$, s               & 1.062$\cdot 10^{-6}$  \\
$\Delta\epsilon_3$        & 374.768 \\
$\tau_3$, s               & 2.162$\cdot 10^{-5}$  \\
$\Delta\epsilon_4$        & 10.387  \\
$\tau_4$, s               & 1.008$\cdot 10^{-8}$  \\ \hline
\end{tabular}
\label{parammeas}
\end{table}

Then, a simulation model corresponding to the measurement case was created. The contents of the model file for the simulation are shown in Fig.~\ref{fig:gprmodel}.

\begin{figure}[!t]
\centering
\fbox{
\begin{minipage}{0.46\textwidth}
\begin{algorithmic}[1]
\State volume (10, 5, 7, 0.125)
\State calctime (15e-6)
\State output (1e-8)
\State abc (cpml, 1.25, 1, 1, 3e-06, 3, 1)
\State block (0, 0, 0, 10, 5, 4, 16.381, 0.002684, deb)
\State debye (deb, 92.103, 7.404e-06, 20.825, 1.062e-06, 374.768, 2.162e-05, 10.387, 1.008e-07)
\State wire (thin, 2.5, 2.5, 3.75, 6.5, 2.5, 3.75, 0.0045)
\State wire (thin, 6.5, 2.5, 3.75, 6.5, 2.5, 4.25, 0.003)
\State wire (thin, 6.5, 2.5, 4.25, 6.75, 2.5, 4.25, 0.003, t)
\State wire (thin, 6.875, 2.5, 4.25, 10, 2.5, 4.25, 0.003)
\State wire (thin, 6.5, 2.5, 4.25, 6.5, 2.75, 4.25, 0.003, t)
\State wire (thin, 6.5, 2.875, 4.25, 6.5, 5, 4.25, 0.003)
\State source (current, 6.75, 2.5, 4.25, 6.875, 2.5, 4.25, 0, fn)
\State calculate (current, 6.75, 2.5, 4.25, 6.875, 2.5, 4.25)
\State calculate (voltage, 6.5, 2.75, 4.25, 6.5, 2.875, 4.25)
\State function (fn, custom, 1e-8, [measured current])
\end{algorithmic}
\end{minipage}}
\caption{Model file for ground potential rise simulation (measured current values are not shown due to limited space).}
\label{fig:gprmodel}
\end{figure}

\begin{figure}[!t]
\centering
\includegraphics[width=3.4in]{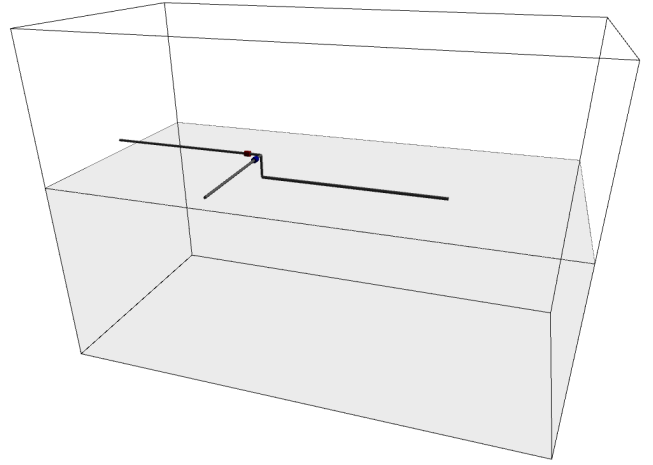}
\caption{Model for calculating ground potential rise.}
\label{fig:wire}
\end{figure}

Thin wires were simulated with the model \cite{railton_treatment_2005}.
Measured current values were set in the "function" command.

\begin{figure}[!t]
\centering
\includegraphics[width=3.7in]{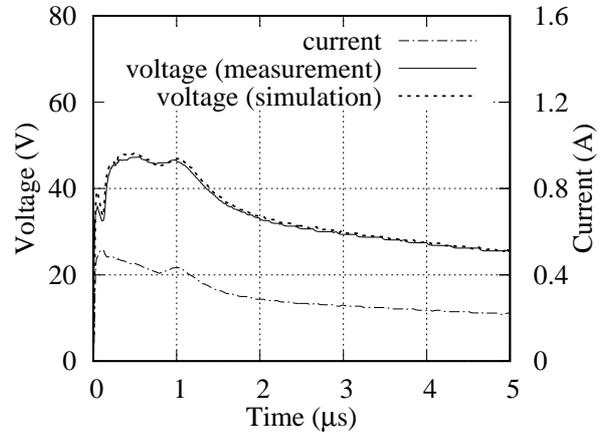}
\caption{Measured and calculated ground potential rise.}
\label{fig:gpr}
\end{figure}

Fig.~\ref{fig:gpr} shows calculated and measured ground potential rise for the model with buried horizontal wire.
The example demonstrates accurate modeling of frequency-dependent soil properties and thin wire models.

\subsection{Lightning overvoltages}
Additionally, the software allows simulating overvoltages caused by lightning strikes. Previously, the code was used for calculations of the back flashover probability \cite{kuklin_choosing_2016}. In that case, the staircase wire model was used for modeling the transmission tower. In the present software version, the oblique wire model \cite{guiffaut_new_2012} was implemented. Therefore, the current software version allows comparing these two wire models.

Below, a simulation example for the direct lightning strike is presented. The tower has the same dimensions as previously \cite{kuklin_choosing_2016}. The tower is modeled with two different wire models: using staircase wires (as in \cite{kuklin_choosing_2016}) and using oblique wires (according to model \cite{guiffaut_new_2012}).

\begin{figure}[!t]
\centering
\includegraphics[width=3.4in]{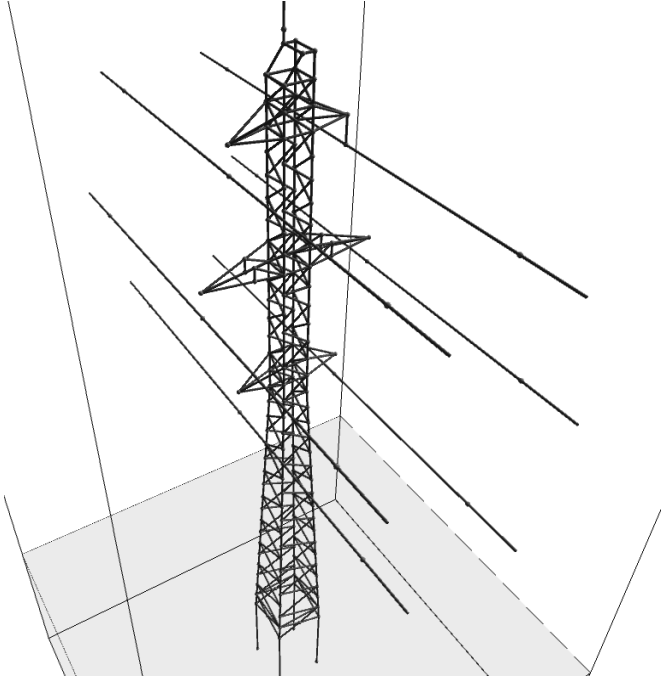}
\caption{Tower model.}
\label{fig:tower}
\end{figure}

Fig.~\ref{fig:tower} shows the 3D view of the simulation model.

\begin{figure}[!t]
\centering
\includegraphics[width=3.7in]{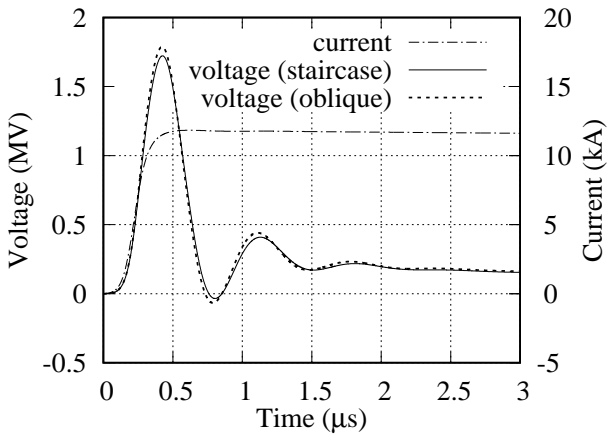}
\caption{Top phase voltage corresponding to simulations with different wire models.}
\label{fig:volt}
\end{figure}

Fig.~\ref{fig:volt} shows the calculated voltage (across top phase insulation). It can be seen that different wire models lead to similar results in this case. Although, the staircase wire model is less accurate for inclined wires \cite{noda_error_2004} (it can be noted, however, that an approach allowing to compensate this error was implemented in the software). Also, the staircase wire model does not allow simulating wires with arbitrary diameters.

\subsection{Calculating apparent soil properties}
The software is useful for conducting simulations related to the apparent soil properties and measurements of those properties. Using the code, it was predicted that for certain array configurations, electromagnetic effects can influence measurement results \cite{kuklin_numerical_2019}. Also, the level of errors was estimated. Later, those effects were confirmed experimentally \cite{kuklin_device_2020}.

An example that calculates apparent soil properties for a three-layer soil model is presented below.

First, Debye parameters corresponding were calculated with the software for soil properties of the three soil layers according to the Messier model. The parameters are presented in Table~\ref{paramcalc}.

\begin{table}[!t]
\renewcommand{\arraystretch}{1.3}
\caption{The $\epsilon_{\infty}$, $\Delta\epsilon$, and $\tau$ parameters of the four-term Debye function expansion.}
\centering
\begin{tabular}{cccc} \hline
$\rho_0$, $\Omega\cdot$m    & 200     & 400     & 800     \\ \hline
$\sigma_0$, mS/m            & 5       & 2.5     & 1.25    \\
$\epsilon_\infty$           & 15.767  & 13.488  & 11.884  \\
$\Delta\epsilon_1$          & 839.441 & 593.525 & 419.744 \\
$\tau_1$, s $\cdot 10^{-5}$ & 3.096   & 3.096   & 3.097   \\
$\Delta\epsilon_2$          & 143.901 & 101.748 & 71.960  \\
$\tau_2$, s $\cdot 10^{-6}$ & 4.068   & 4.066   & 4.069   \\
$\Delta\epsilon_3$          & 66.971  & 47.348  & 33.494  \\
$\tau_3$, s $\cdot 10^{-7}$ & 7.217   & 7.212   & 7.218   \\
$\Delta\epsilon_4$          & 31.515  & 22.273  & 15.757  \\
$\tau_4$, s $\cdot 10^{-8}$ & 9.141   & 9.132   & 9.140   \\ \hline
\end{tabular}
\label{paramcalc}
\end{table}

\begin{figure}[!t]
\centering
\fbox{
\begin{minipage}{0.46\textwidth}
\begin{algorithmic}[1]
\State volume (5, 5, 4, 0.05)
\State calctime (400e-6)
\State output (1e-008)
\State abc (cpml, 0.5, 1, 1, 1e-3, 3, 1)
\State block (0, 0, 2, 5, 5, 3, 15.767, 0.005, deb)
\State debye (deb, 839.441, 3.096e-05, 143.901, 4.068e-06, 66.971, 7.217e-07, 31.515, 9.141e-08)
\State block (0, 0, 1, 5, 5, 2, 13.488, 0.0025, deb1)
\State debye (deb1, 593.525, 3.096e-05, 101.748, 4.066e-06, 47.348, 7.212e-07, 22.273, 9.132e-08)
\State block (0, 0, 0, 5, 5, 1, 11.884, 0.00125, deb2)
\State debye (deb2, 419.744, 3.097e-05, 71.960, 4.069e-06, 33.494, 7.218e-07, 15.757, 9.140e-08)
\State wire (thin, 1, 1, 2.95, 1, 1, 3.15, 0.008)
\State wire (thin, 2, 1, 2.95, 2, 1, 3.15, 0.008)
\State wire (thin, 1, 1, 3.15, 1.45, 1, 3.15, 0.003, t)
\State wire (thin, 1.5, 1, 3.15, 2, 1, 3.15, 0.003)
\State wire (thin, 4, 3, 2.95, 4, 3, 3.15, 0.008)
\State wire (thin, 4, 4, 2.95, 4, 4, 3.15, 0.008)
\State wire (thin, 4, 3, 3.15, 4, 3.45, 3.15, 0.003, t)
\State wire (thin, 4, 3.5, 3.15, 4, 4, 3.15, 0.003)
\State source (current, 1.45, 1, 3.15, 1.5, 1, 3.15, 0, fn)
\State calculate (current, 1.45, 1, 3.15, 1.5, 1, 3.15)
\State calculate (voltage, 4, 3.45, 3.15, 4, 3.5, 3.15)
\State function (fn, heidler, s, 3.7e-7, 1.4e-5, 1, 10, 1, 1.85e-7)
\end{algorithmic}
\end{minipage}}
\caption{Model file for array simulation.}
\label{fig:arraymodel}
\end{figure}

\begin{figure}[!t]
\centering
\includegraphics[width=3.4in]{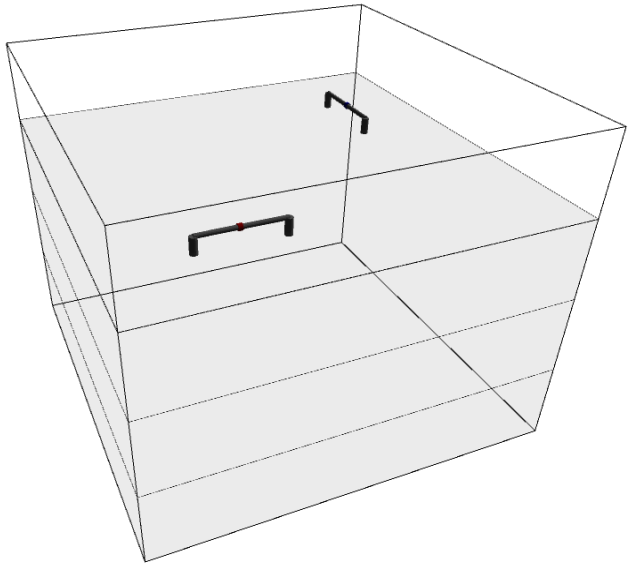}
\caption{Array model.}
\label{fig:array}
\end{figure}

Then, the simulation model was created. Listing for the model file is presented in Fig.~\ref{fig:arraymodel}. 3D view of the array model is shown in Fig.~\ref{fig:array}.

After FDTD simulations, apparent soil properties were calculated from the known current and voltage. The software can calculate the properties automatically if the dimensions of the array are set (in the corresponding CLI command).

Additionally, apparent resistivity was calculated with the method \cite{takahashi_analysis_1990} (implemented in the software).

\begin{figure}[!t]
\centering
\includegraphics[width=3.7in]{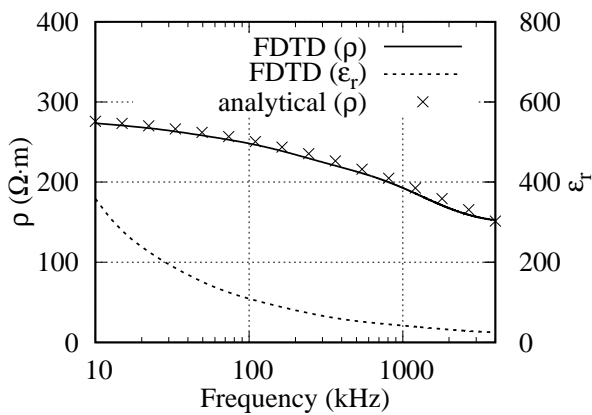}
\caption{Apparent soil properties calculated with different methods.}
\label{fig:prop}
\end{figure}

Apparent soil properties calculated with two different methods are shown in Fig.~\ref{fig:prop}. It can be seen that both methods give the same resistivity results.

\section{Conclusion}
The open-source software for electrical engineering applications was presented in the work.

The software allows conducting simulations with complicated structures consisting of thin wires, and accurate modeling of frequency-dependent soil properties. It provides various additional numerical methods for lightning-related simulations and tools for measuring soil properties.
In other words, it provides numerical methods not available in other open-source FDTD software.

It has been demonstrated that the software can be successfully used for calculations of electrical grounding parameters, overvoltages, apparent resistivity and permittivity values, etc.

The work presented several simulation cases that can be conducted using CLI. However, the same simulations can be performed using the C++ library. This allows simulations to be performed at a lower level: for example, when modifying numerical methods is required.

The convenient structure of the software makes further development of the software modules relatively independent.
Thus, many new features can be added in the future.

\bibliography{references}

\begin{thebibliography}{10}
\expandafter\ifx\csname url\endcsname\relax
  \def\url#1{\texttt{#1}}\fi
\expandafter\ifx\csname urlprefix\endcsname\relax\def\urlprefix{URL }\fi
\expandafter\ifx\csname href\endcsname\relax
  \def\href#1#2{#2} \def\path#1{#1}\fi

\bibitem{grcev_electromagnetic_1990}
L.~D. Grcev, F.~Dawalibi, An electromagnetic model for transients in grounding
  systems, IEEE Transactions on Power Delivery 5~(4) (1990) 1773--1781.

\bibitem{tanabe_novel_2001}
K.~Tanabe, Novel method for analyzing the transient behavior of grounding
  systems based on the finite-difference time-domain method, in: {IEEE} {Power}
  {Engineering} {Society} {Winter} {Meeting}, 2001, Vol.~3, 2001, pp.
  1128--1132.
\newblock \href {http://dx.doi.org/10.1109/PESW.2001.917230}
  {\path{doi:10.1109/PESW.2001.917230}}.

\bibitem{yutthagowith_application_2011}
P.~Yutthagowith, A.~Ametani, N.~Nagaoka, Y.~Baba, Application of the partial
  element equivalent circuit method to tower surge response calculations, IEEJ
  Transactions On Electrical And Electronic Engineering 6~(4) (2011) 324--330.

\bibitem{olsen_comparison_1996}
R.~G. Olsen, M.~C. Willis, A comparison of exact and quasi-static methods for
  evaluating grounding systems at high frequencies, IEEE Transactions on Power
  Delivery 11~(2) (1996) 1071--1081.
\newblock \href {http://dx.doi.org/10.1109/61.489370}
  {\path{doi:10.1109/61.489370}}.

\bibitem{kuklin_choosing_2016}
D.~Kuklin, Choosing configurations of transmission line tower grounding by back
  flashover probability value, Frontiers in Energy 10~(2) (2016) 213--226.
\newblock \href {http://dx.doi.org/10.1007/s11708-016-0398-6}
  {\path{doi:10.1007/s11708-016-0398-6}}.

\bibitem{kuklin_extension_2016}
D.~Kuklin, Extension of {Thin} {Wire} {Techniques} in the {FDTD} {Method} for
  {Debye} {Media}, Progress In Electromagnetics Research M 51 (2016) 9--17.
\newblock \href {http://dx.doi.org/10.2528/PIERM16081804}
  {\path{doi:10.2528/PIERM16081804}}.

\bibitem{kuklin_numerical_2019}
D.~Kuklin, Numerical {Analysis} of {Electromagnetic} {Coupling} {Effects} in
  {Measurements} of {Frequency} {Dependent} {Soil} {Electrical} {Properties},
  Progress In Electromagnetics Research M 79 (2019) 101--111.
\newblock \href {http://dx.doi.org/10.2528/PIERM18112102}
  {\path{doi:10.2528/PIERM18112102}}.

\bibitem{kuklin_device_2020}
D.~Kuklin, Device for the field measurements of frequency-dependent soil
  properties in the frequency range of lightning currents, Review of Scientific
  Instruments 91~(11) (2020) 114701, publisher: American Institute of Physics.
\newblock \href {http://dx.doi.org/10.1063/5.0012126}
  {\path{doi:10.1063/5.0012126}}.

\bibitem{warren_gprmax_2016}
C.~Warren, A.~Giannopoulos, I.~Giannakis, {gprMax}: {Open} source software to
  simulate electromagnetic wave propagation for {Ground} {Penetrating} {Radar},
  Computer Physics Communications 209 (2016) 163--170.
\newblock \href {http://dx.doi.org/10.1016/j.cpc.2016.08.020}
  {\path{doi:10.1016/j.cpc.2016.08.020}}.

\bibitem{noauthor_meep_2022}
{Meep}. {Meep} is a free and open-source software package for {electromagnetics}
  {simulation}.
\newline\urlprefix\url{https://github.com/NanoComp/meep}

\bibitem{yee_numerical_1966}
K.~S. Yee, Numerical solution of initial boundary value problems involving
  {Maxwell}'s equations in isotropic media, IEEE Transactions on Antennas and
  Propagation 14~(3) (1966) 302--307.

\bibitem{taflove_numerical_1975}
A.~Taflove, M.~E. Brodwin, Numerical solution of steady-state electromagnetic
  scattering problems using the time-dependent {Maxwell}'s equations, IEEE
  Transactions on Microwave Theory and Techniques 23~(8) (1975) 623--630.

\bibitem{taflove_computational_2005}
A.~Taflove, S.~C. Hagness, Computational electrodynamics: the finite-difference
  time-domain method, 3rd Edition, Artech House, 2005.

\bibitem{berenger_perfectly_1994}
J.~P. Berenger, A perfectly matched layer for the absorption of electromagnetic
  waves, Journal of Computational Physics 114~(2) (1994) 185--200.

\bibitem{roden_convolution_2000}
J.~A. Roden, S.~D. Gedney, Convolution {PML} ({CPML}): {An} efficient {FDTD}
  implementation of the {CFS}–{PML} for arbitrary media, Microwave and
  Optical Technology Letters 27~(5) (2000) 334--339.
\newblock \href
  {http://dx.doi.org/10.1002/1098-2760(20001205)27:5<334::AID-MOP14>3.0.CO;2-A}
  {\path{doi:10.1002/1098-2760(20001205)27:5<334::AID-MOP14>3.0.CO;2-A}}.

\bibitem{railton_treatment_2005}
C.~J. Railton, D.~L. Paul, I.~J. Craddock, G.~S. Hilton, The treatment of
  geometrically small structures in {FDTD} by the modification of assigned
  material parameters, IEEE Transactions on Antennas and Propagation 53~(12)
  (2005) 4129--4136.

\bibitem{taniguchi_improved_2008}
Y.~Taniguchi, Y.~Baba, N.~Nagaoka, A.~Ametani, An {Improved} {Thin} {Wire}
  {Representation} for {FDTD} {Computations}, IEEE Transactions on Antennas and
  Propagation 56~(10) (2008) 3248--3252, conference Name: IEEE Transactions on
  Antennas and Propagation.
\newblock \href {http://dx.doi.org/10.1109/TAP.2008.929447}
  {\path{doi:10.1109/TAP.2008.929447}}.

\bibitem{taniguchi_improved_2009}
Y.~Taniguchi, Y.~Baba, N.~Nagaoka, A.~Ametani, An improved
  arbitrary-radius-wire representation for {FDTD} electromagnetic and surge
  calculations, Kyoto, Japan, 2009.

\bibitem{montoya_modeling_1999}
T.~P. Montoya, J.~G. Maloney, G.~S. Smith, Modeling staircased wires using the
  {FDTD} method, IEEE Antennas and Propagation Society International Symposium
  1 (1999) 180--183.

\bibitem{noda_error_2004}
T.~Noda, R.~Yonezawa, S.~Yokoyama, Y.~Takahashi, Error in propagation velocity
  due to staircase approximation of an inclined thin wire in {FDTD} surge
  simulation, IEEE Transactions on Power Delivery 19~(4) (2004) 1913--1918.

\bibitem{guiffaut_new_2012}
C.~Guiffaut, A.~Reineix, B.~Pecqueux, New oblique thin wire formalism in the
  {FDTD} method with multiwire junctions, IEEE Transactions on Antennas and
  Propagation 60~(3) (2012) 1458--1466.

\bibitem{akbari_effect_2013}
M.~Akbari, K.~Sheshyekani, M.~R. Alemi, The {Effect} of {Frequency}
  {Dependence} of {Soil} {Electrical} {Parameters} on the {Lightning}
  {Performance} of {Grounding} {Systems}, IEEE Transactions on Electromagnetic
  Compatibility 55~(4) (2013) 739--746.
\newblock \href {http://dx.doi.org/10.1109/TEMC.2012.2222416}
  {\path{doi:10.1109/TEMC.2012.2222416}}.

\bibitem{cavka_comparison_2014}
D.~Cavka, N.~Mora, F.~Rachidi, A {Comparison} of {Frequency}-{Dependent} {Soil}
  {Models}: {Application} to the {Analysis} of {Grounding} {Systems}, IEEE
  Transactions on Electromagnetic Compatibility 56~(1) (2014) 177--187.
\newblock \href {http://dx.doi.org/10.1109/TEMC.2013.2271913}
  {\path{doi:10.1109/TEMC.2013.2271913}}.

\bibitem{alipio_frequency_2013}
R.~Alipio, S.~Visacro, Frequency dependence of soil parameters: effect on the
  lightning response of grounding electrodes, IEEE Transactions on
  Electromagnetic Compatibility 55~(1).
\newblock \href {http://dx.doi.org/10.1109/TEMC.2012.2210227}
  {\path{doi:10.1109/TEMC.2012.2210227}}.

\bibitem{datsios_methods_2020}
Z.~G. Datsios, P.~N. Mikropoulos, E.~T. Staikos, Methods for {Field}
  {Measurement} of the {Frequency}-{Dependent} {Soil} {Electrical}
  {Properties}: {Evaluation} of {Electrode} {Arrangements} {Through} {FEM}
  {Computations}, in: B.~Németh (Ed.), Proceedings of the 21st {International}
  {Symposium} on {High} {Voltage} {Engineering}, Lecture {Notes} in
  {Electrical} {Engineering}, Springer International Publishing, Cham, 2020,
  pp. 1328--1339.
\newblock \href {http://dx.doi.org/10.1007/978-3-030-31676-1_124}
  {\path{doi:10.1007/978-3-030-31676-1_124}}.

\bibitem{visacro_lightning_2015}
S.~Visacro, R.~Alipio, C.~Pereira, M.~Guimarães, M.~A.~O. Schroeder, Lightning
  {Response} of {Grounding} {Grids}: {Simulated} and {Experimental} {Results},
  IEEE Transactions on Electromagnetic Compatibility 57~(1) (2015) 121--127.
\newblock \href {http://dx.doi.org/10.1109/TEMC.2014.2362091}
  {\path{doi:10.1109/TEMC.2014.2362091}}.

\bibitem{li_inversion_2020}
Z.-X. Li, S.-W. Rao, The {Inversion} of {One}-{Dimensional} {Soil} {Parameters}
  in the {Frequency} {Domain} {With} {Considering} {Multilayered} {Earth}
  {Based} on {Simulated} {Annealing} {Algorithm}, IEEE Transactions on
  Electromagnetic Compatibility 62~(2) (2020) 425--432.
\newblock \href {http://dx.doi.org/10.1109/TEMC.2019.2906486}
  {\path{doi:10.1109/TEMC.2019.2906486}}.

\bibitem{okoniewski_simple_1997}
M.~Okoniewski, M.~Mrozowski, M.~A. Stuchly, Simple treatment of multi-term
  dispersion in {FDTD}, IEEE Microwave and Guided Wave Letters 7~(5) (1997)
  121--123.
\newblock \href {http://dx.doi.org/10.1109/75.569723}
  {\path{doi:10.1109/75.569723}}.

\bibitem{kelley_debye_2007}
D.~F. Kelley, T.~J. Destan, R.~J. Luebbers, Debye function expansions of
  complex permittivity using a hybrid particle swarm-least squares optimization
  approach, IEEE Transactions on Antennas and Propagation 55~(7) (2007)
  1999--2005.
\newblock \href {http://dx.doi.org/10.1109/TAP.2007.900230}
  {\path{doi:10.1109/TAP.2007.900230}}.

\bibitem{takahashi_analysis_1990}
T.~Takahashi, T.~Kawase, Analysis of apparent resistivity in a multi-layer
  earth structure, IEEE Transactions on Power Delivery 5~(2) (1990) 604--612,
  conference Name: IEEE Transactions on Power Delivery.
\newblock \href {http://dx.doi.org/10.1109/61.53062}
  {\path{doi:10.1109/61.53062}}.

\bibitem{loke_tutorial_2020}
M.~H. Loke, Tutorial: 2-{D} and 3-{D} electrical imaging surveys, 2020.

\bibitem{roy_depth_1971}
A.~Roy, A.~Apparao, Depth of investigation in direct current methods,
  GEOPHYSICS 36~(5) (1971) 943--959.
\newblock \href {http://dx.doi.org/10.1190/1.1440226}
  {\path{doi:10.1190/1.1440226}}.

\bibitem{barker_depth_1989}
R.~Barker, Depth of investigation of collinear symmetrical four‐electrode
  arrays, GEOPHYSICS 54~(8) (1989) 1031--1037.
\newblock \href {http://dx.doi.org/10.1190/1.1442728}
  {\path{doi:10.1190/1.1442728}}.

\bibitem{alipio_modeling_2014}
R.~Alipio, S.~Visacro, Modeling the {Frequency} {Dependence} of {Electrical}
  {Parameters} of {Soil}, IEEE Transactions on Electromagnetic Compatibility
  56~(5) (2014) 1163--1171.
\newblock \href {http://dx.doi.org/10.1109/TEMC.2014.2313977}
  {\path{doi:10.1109/TEMC.2014.2313977}}.

\bibitem{portela_measurement_1999}
C.~Portela, Measurement and modeling of soil electromagnetic behavior, in: 1999
  {IEEE} {International} {Symposium} on {Electromagnetic} {Compatability}.
  {Symposium} {Record} ({Cat}. {No}.{99CH36261}), Vol.~2, 1999, pp. 1004--1009
  vol.2.
\newblock \href {http://dx.doi.org/10.1109/ISEMC.1999.810203}
  {\path{doi:10.1109/ISEMC.1999.810203}}.

\bibitem{shindo_new_1985}
T.~Shindo, T.~Suzuki, A {New} {Calculation} {Method} of {Breakdown}
  {Voltage}-{Time} {Characteristics} of {Long} {Air} {Gaps}, IEEE Transactions
  on Power Apparatus and Systems PAS-104~(6) (1985) 1556--1563, conference
  Name: IEEE Transactions on Power Apparatus and Systems.
\newblock \href {http://dx.doi.org/10.1109/TPAS.1985.319172}
  {\path{doi:10.1109/TPAS.1985.319172}}.

\bibitem{pigini_performance_1989}
A.~Pigini, G.~Rizzi, E.~Garbagnati, A.~Porrino, G.~Baldo, G.~Pesavento,
  Performance of large air gaps under lightning overvoltages: experimental
  study and analysis of accuracy predetermination methods, IEEE Transactions on
  Power Delivery 4~(2) (1989) 1379--1392.
\newblock \href {http://dx.doi.org/10.1109/61.25625}
  {\path{doi:10.1109/61.25625}}.

\bibitem{chisholm_new_2010}
W.~A. Chisholm, New challenges in lightning impulse flashover modeling of air
  gaps and insulators, IEEE Electrical Insulation Magazine 26~(2) (2010)
  14--25.
\newblock \href {http://dx.doi.org/10.1109/MEI.2010.5482551}
  {\path{doi:10.1109/MEI.2010.5482551}}.

\bibitem{heidler_class_2002}
F.~Heidler, J.~Cvetic, A class of analytical functions to study the lightning
  effects associated with the current front, European Transactions on
  Electrical Power 12~(2) (2002) 141--150.

\bibitem{noauthor_guide_1991}
Guide to procedures for estimating the lightning performance of transmission
  lines, Cigre, 1991.

\bibitem{noauthor_qt_nodate}
\href{https://www.qt.io}{Qt. {Cross}-platform software development for embedded
  \& desktop}.
\newline\urlprefix\url{https://www.qt.io}

\bibitem{noauthor_opengl_nodate}
\href{https://www.opengl.org/}{{OpenGL} - {The} {Industry} {Standard} for
  {High} {Performance} {Graphics}}.
\newline\urlprefix\url{https://www.opengl.org/}

\end{thebibliography}

\end{document}